\documentclass[letterpaper]{article} 
\usepackage{aaai2026}  
\usepackage{times}  
\usepackage{helvet}  
\usepackage{courier}  
\usepackage[hyphens]{url}  
\usepackage{graphicx} 
\usepackage{natbib}  
\usepackage{caption} 
\usepackage{algorithm}
\usepackage{algorithmic}

\usepackage{newfloat}
\usepackage{listings}
\DeclareCaptionStyle{ruled}{labelfont=normalfont,labelsep=colon,strut=off} 
\floatstyle{ruled}
\newfloat{listing}{tb}{lst}{}
\floatname{listing}{Listing}
\usepackage{color}
\usepackage[cmyk]{xcolor}
\usepackage{amsmath}
\usepackage{amsfonts} 
\usepackage{booktabs}
\usepackage{tabularx}
\usepackage{multirow}

\title{Hard vs. Noise: Resolving Hard-Noisy Sample Confusion in Recommender Systems via Large Language Models}
\author {
    Tianrui Song\textsuperscript{\rm 1}, 
    Wen-Shuo Chao\textsuperscript{\rm 1},
    Hao Liu\textsuperscript{\rm 1,\rm 2}\thanks{Corresponding author.}
}
\affiliations {
    \textsuperscript{\rm 1}The Hong Kong University of Science and Technology (Guangzhou)\\
    \textsuperscript{\rm 2}The Hong Kong University of Science and Technology\\
    tsong847@connect.hkust-gz.edu.cn, wschao829@connect.hkust-gz.edu.cn, liuh@ust.hk
}
\usepackage{bibentry}

\begin{document}

\maketitle

\begin{abstract}
Implicit feedback, employed in training recommender systems, unavoidably confronts noise due to factors such as misclicks and position bias.
Previous studies have attempted to identify noisy samples through their diverged data patterns, such as higher loss values, and mitigate their influence through sample dropping or reweighting.
However, we observed that noisy samples and hard samples display similar patterns, leading to hard-noisy confusion issue.
Such confusion is problematic as hard samples are vital for modeling user preferences.
To solve this problem, we propose LLMHNI framework, leveraging two auxiliary user-item relevance signals generated by Large Language Models (LLMs) to differentiate hard and noisy samples.
LLMHNI obtains user-item semantic relevance from LLM-encoded embeddings, which is used in negative sampling to select hard negatives while filtering out noisy false negatives.
An objective alignment strategy is proposed to project LLM-encoded embeddings, originally for general language tasks, into a representation space optimized for user-item relevance modeling.
LLMHNI also exploits LLM-inferred logical relevance within user-item interactions to identify hard and noisy samples.
These LLM-inferred interactions are integrated into the interaction graph and guide denoising with cross-graph contrastive alignment.
To eliminate the impact of unreliable interactions induced by LLM hallucination, we propose a graph contrastive learning strategy that aligns representations from randomly edge-dropped views to suppress unreliable edges.
Empirical results demonstrate that LLMHNI significantly improves denoising and recommendation performance.
\end{abstract}

\begin{links}
    \link{Code}{https://github.com/TianRui-Song717/LLMHNI}
\end{links}

\section{Introduction}
Recommender Systems (RS) rely on implicit feedback, such as clicks and purchases, to model user preferences ~\cite{He2020LightGCN,Luo2020SpatialOR}.
Traditionally, these interactions are labeled positively if observed and negatively if not ~\cite{Ding2020Simplify,Wang2021DenoisingIF}.
However, this schema is questioned due to the false-positive noise from misclicks and false-negative noise from position bias ~\cite{Wang2021ClicksCB}.
To address such noise issues, denoising recommendation strategies have emerged, including sample dropping and sample reweighting.
Sample dropping mitigates the impact of noise by removing noisy interactions during training ~\cite{Chen2021GeneralizedDW}, while reweighting assigns lower weights to noisy interactions ~\cite{Wang2022EfficientBO,Gao2022SelfGuidedLT}.
These techniques hinge on accurately distinguishing between clean and noisy samples by their divergent patterns in loss value ~\cite{Ding2020Simplify}, prediction scores ~\cite{Wang2021DenoisingIF}, and gradients ~\cite{Wang2022EfficientBO}.

\begin{figure}[t]
    \centering
    \includegraphics[width=\linewidth]{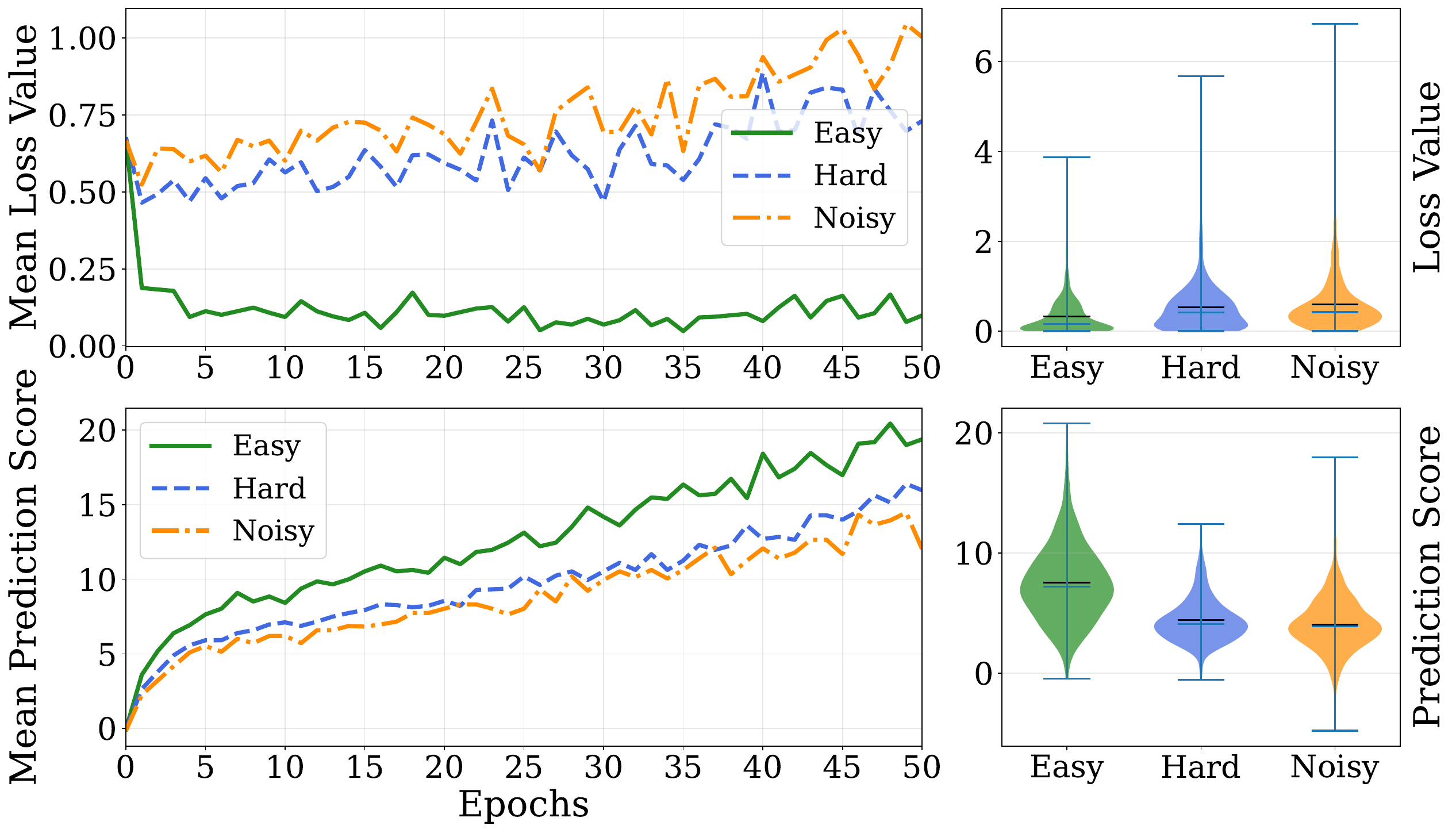}
    \caption{On the left, we demonstrate that hard and noisy samples display similar patterns in both loss values and prediction scores throughout the training process. On the right, we take the results from the 5th epoch as an example to illustrate how the prediction scores and loss values of hard and noisy samples overlap in distribution. Additional details about this figure can be found in the appendix.}
    \label{fig:Introduction}
\end{figure}

Despite their advancements, these denoising methods often face the challenge of misidentifying hard samples as noisy ones.
As illustrated in Figure~\ref{fig:Introduction}, while noisy samples exhibit distinct patterns compared to easy samples, we observed that hard samples and noisy samples tend to present similar patterns in both prediction scores and loss values.
Consequently, previous denoising approaches that rely solely on data patterns \emph{struggle to distinguish between hard and noisy samples}.
This misidentification is problematic because hard samples have been shown to be beneficial, both empirically~\cite{Gantner2012PersonalizedRF} and theoretically~\cite{Shi2023OnTT}.
Mistakenly treating hard samples as noise during training ultimately leads to suboptimal results.

\begin{figure}[t]
    \centering
    \includegraphics[width=\linewidth]{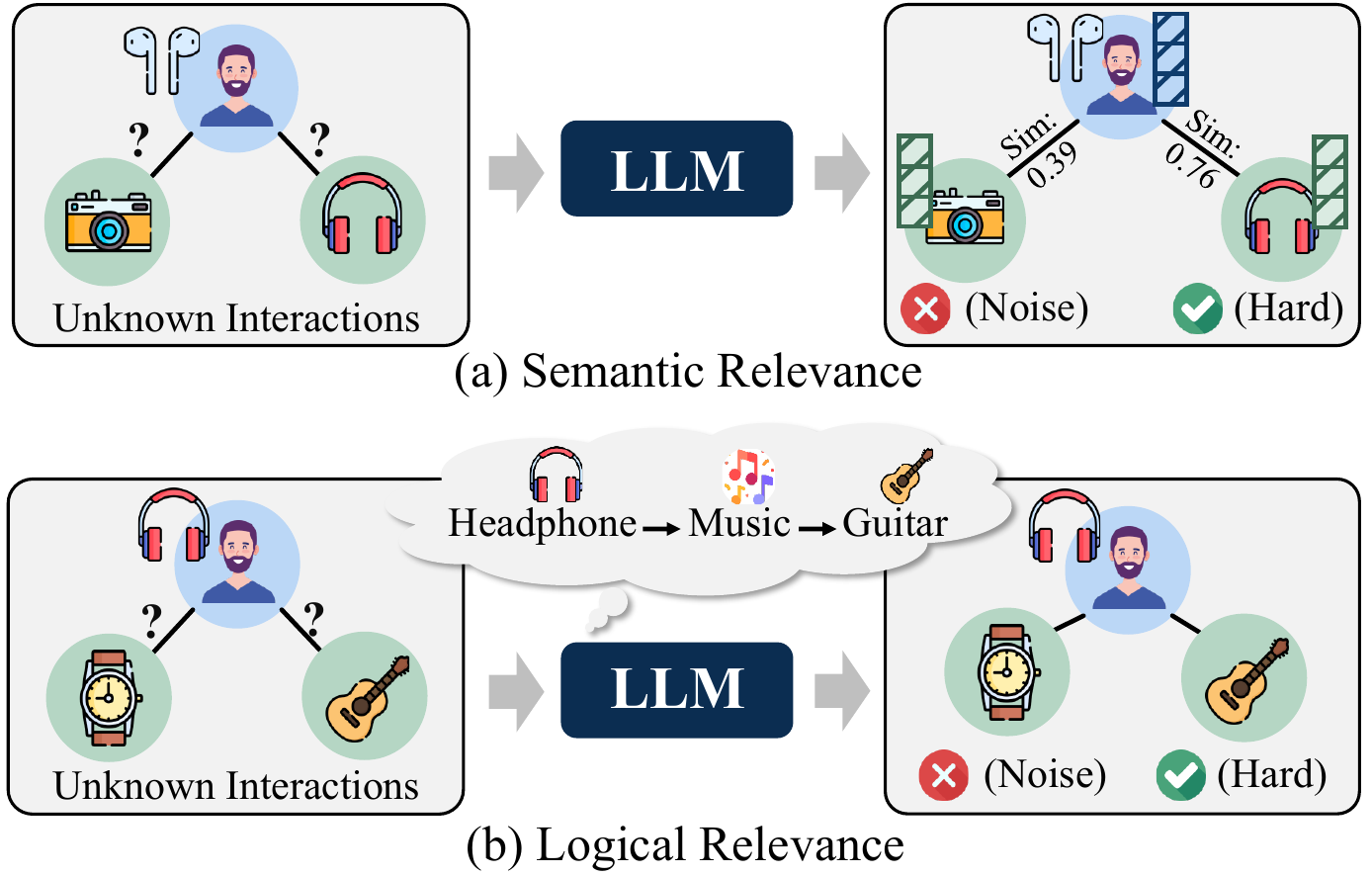}
    \caption{Semantic Relevance and Logical Relevance.}
    \label{fig:IntroRelevance}
\end{figure}

Since distinguishing hard and noisy samples based solely on numerical patterns derived from user-item collaborative information is insufficient, addressing this issue necessitates auxiliary signals.
Recently, Large Language Models (LLMs) have emerged as powerful tools to enhance recommender systems.
Existing approaches take the knowledge generated by LLMs as supplementary information beyond collaborative signals in recommender systems \cite{lin2025can}.
Inspired by their promising performance, we employ LLMs to provide auxiliary information for distinguishing hard samples from noisy ones.
Specifically, we exploit two types of user-item relevance signals that 
distinct from those captured by user-item interactions.
1) Semantic Relevance in LLM-encoded embedding:
As shown in Fig. \ref{fig:IntroRelevance}(a), LLM-encoded user and item embeddings offer semantic relevance between users and items, which helps identify hard and noisy samples with relevance scores.
2) Logical Relevance in LLM-inferred interactions:
LLMs possess reasoning capabilities that can infer logical relevance within user-item interactions and distinguish hard samples from noisy ones.
As shown in Fig. \ref{fig:IntroRelevance}(b), LLM deduces that a user bought headphones enjoys music and might therefore be interested in a guitar.

However, leveraging these two auxiliary relevance signals to distinguish hard and noisy samples in recommender systems faces two challenges:
1) \textbf{Objective-Mismatched Embeddings}: LLM-encoded embeddings, trained for general language tasks rather than user preference modeling, suffer an objective mismatch for recommendation tasks.
Consequently, the user-item similarity values derived from these objective-mismatched embeddings can mislead the identification between hard and noisy samples and hinder recommendation model performance.
2) \textbf{Hallucination-Induced Interactions}: LLMs suffer from hallucination, which undermines the reliability of their inferred user-item interactions.
Including these hallucination-induced interactions during training may amplify label noise and propagate hallucination errors into the recommendation model.

To overcome aforementioned challenges, we introduce the \textbf{L}arge \textbf{L}anguage \textbf{M}odels enhanced \textbf{H}ard-\textbf{N}oisy sample \textbf{I}dentification framework (\textbf{LLMHNI}).
It comprises two modules that take auxiliary signals generated by LLMs to differentiate hard and noisy samples, improving the denoising process.
The first module, Semantic Relevance Guided Hard Negative Mining, harnesses LLMs to encode text profiles of users and items.
Semantic relevance (i.e., embedding similarities) between users and items are used to guide negative sampling, facilitating the selection of hard negatives while avoiding the introduction of false negatives.
To further mitigate the \textit{objective-mismatched embedding}, we design an objective alignment strategy that projects raw LLM-encoded embeddings into a tailored representation space optimized for preference modeling.
The second module, Logical Relevance Guided Interaction Denoising, employs LLMs to infer logical relevance within user-item interactions, identifying hard and noisy ones.
These interactions are integrated into interaction graph and guide interaction denoise.
Specifically, we design a cross-graph contrastive alignment that suppresses interactions inconsistent between the original graph and the one enhanced with LLM-inferred hard and noisy interactions.
To mitigate \textit{hallucination-induced interactions} within interaction graph, a graph contrastive learning strategy is incorporated, which suppresses hallucination-induced edges by aligning representations from two randomly edge-dropped views of the interaction graph.

Our main contributions are summarized as follows.
\begin{itemize}
    \item We propose \textbf{LLMHNI}, a novel framework that takes semantic relevance signals in LLM-encoded embeddings and logical relevance signals in LLM-inferred interactions to guide negative sampling and interaction denoising, addressing the noisy-hard sample confusion in RS.
    \item LLMHNI addresses the objective mismatch of LLM-encoded embeddings by projecting the raw embedding into an aligned representation space. It also reduces the influence of hallucination-induced interactions inferred by LLM with a graph contrastive learning strategy.
    \item Extensive experiments on three real-world datasets and two backbone recommenders demonstrate the effectiveness of our method. Results show that LLMHNI delivers impressive performance and robust noise resilience. 
\end{itemize}

\section{Preliminary}
The objective of training a recommender system is to learn a scoring function $\hat{y}_{u,i} = f_{\theta}(u,i)$ from interactions between users $u \in \mathcal{U}$ and items $i \in \mathcal{I}$. 
We assume that user-interacted items ${y}^{*}_{u,i}= 1$ are preferred by the user, while those not interacted $y^{*}_{u,i}= 0$ are not.
To optimize the scoring function $f_{\theta}(u,i)$, we employ Bayesian Personalized Ranking (BPR) loss as loss function $\mathcal{L}_{rec}$, which are formulated as follows:
\begin{align}
\label{eq:bpr}
\mathcal{L}_{\mathrm{BPR}}(\mathcal{D}^*)
= \mathop{-\mathbb{E}[\log(\sigma}\limits_{(u,i,j)\sim \mathbf{P}_{\mathcal{D}^{*}}}(\hat y_{u,i}-\hat y_{u,j}))],
\end{align}
where $j$ denotes negative items sampled according to the distribution $\mathbf{P}_{\mathcal{D}^{*}}$, and $\mathcal{D}^{*}= \{(u,i, y_{u,i}^{*}) \mid u \in \mathcal{U}, i \in \mathcal{I}\}$ represents the dataset.
$\sigma$ denotes the sigmoid.
The optimal parameter $\theta^{*}$ is obtained by minimizing the $\mathcal{L}_{rec}$:
\begin{align}
    \label{eq:recobj}
    \theta^{*} = \mathop{\arg\min}\limits_{\theta}\mathcal{L}_{rec}(\mathcal{D}^{*}),
\end{align}
But this assumption is unreliable for two reasons: 
\emph{(1) False positive issue}, user-interacted items might not reflect real user preference due to factors such as accidental clicks and position bias.
\emph{(2) False negative issue}, non-interacted items are not necessarily user dislikes, they may have been overlooked due to factors such as suboptimal display positions.
These issues introduce noisy interactions, formally defined as $\tilde{\mathcal{D}} = \{(u,i,\tilde{y}) \mid \tilde{y} \neq y^{*}\}$. To address this, in this work, we formulate \emph{denoising recommender training task} as:
\begin{equation}
    \label{eq:noiseobj}
    \theta^{*} = \mathop{\arg\min}\limits_{\theta}\mathcal{L}_{rec}(\mathcal{D}^{*} \cup \tilde{\mathcal{D}}, \mathcal{P}_{\mathcal{U}}, \mathcal{P}_{\mathcal{I}}),
\end{equation}
where $\mathcal{P}_{\mathcal{U}} = \{\mathcal{P}_{u} | u \in \mathcal{U}\}, \mathcal{P}_{\mathcal{I}} = \{\mathcal{P}_{i} | i \in \mathcal{I}\}$
are the text profiles of users and items that describe user prefereces and item characteristics, respectively.

\section{Proposed Method}

\begin{figure*}[htbp!]
  \centering
  \includegraphics[width=\textwidth]{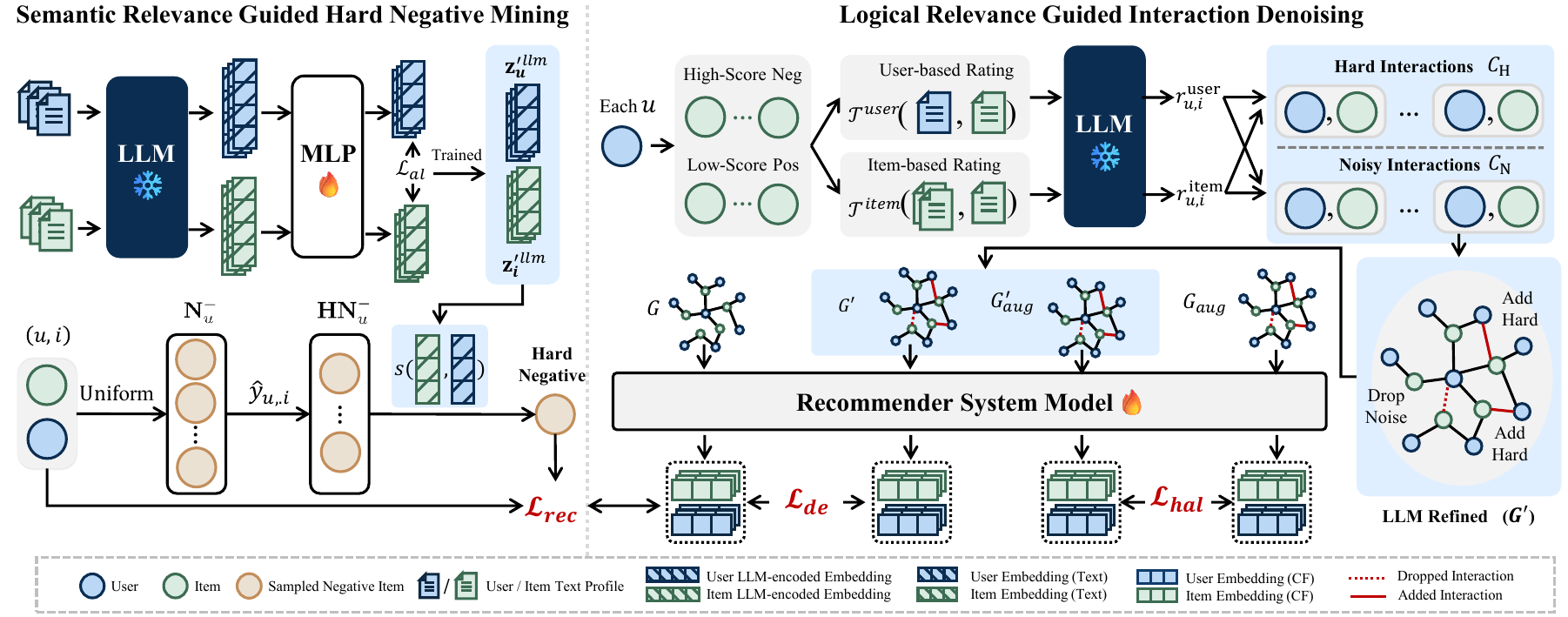}
  \caption{The overview of our proposed LLMHNI framework.}
  \label{fig:overview}
\end{figure*}

We present the \textbf{LLMHNI}, a novel framework that harnesses LLM-generated auxiliary signals to resolve the \textit{hard-noisy sample confusion}.
As illustrated in Fig. \ref{fig:overview}, LLMHNI integrates two modules:
(1) \textit{Semantic Relevance Guided Hard Negative Mining} leverages user and item embeddings encoded by LLMs to capture semantic relevance, facilitating effective sampling of hard negatives while minimizing the risk of false negatives.
(2) \textit{Logical Relevance Guided Interaction Denoising} employs LLM-inferred logical user-item relevance to identify hard and noisy interactions, thereby refining the interaction graph and samples for noise mitigation.

\subsection{Semantic Relevance Guided Hard Negative Mining}
LLM-encoded textual embeddings inherently capture semantic relevance between users and items, which are critical auxiliary information benefiting hard-noisy sample differentiation.
In this part, we leverage these inherent semantic relevance signals to select hard negative items while filtering out false negatives for recommender system.

\subsubsection{Objective-Aligned Embedding Generation.}
User and item text embeddings encoded by LLMs reflect user-item semantic relevance.
However, LLMs are trained for language modeling, which hinders the embeddings' effectiveness in reflecting the user-item correlation required for the recommendation.
To address this issue, we project embeddings to an optimized representation space.
We utilize the LLM embedding model $\text{LLM}_{\text{enc}}$ to encode the text profiles of user ($\mathcal{P}_{u}$) and item ($\mathcal{P}_{i}$).
An MLP projects embeddings to a low-dimensional representation space for objective alignment
\begin{equation}
    \mathbf{e}^{llm}_{u} = \text{LLM}_{\text{enc}}(\mathcal{P}_{u}), \quad
    \mathbf{e}^{llm}_{i} = \text{LLM}_{\text{enc}}(\mathcal{P}_{i}),
\end{equation}
\begin{equation}
    \mathbf{z}_{u}^{llm} = \text{MLP}(\mathbf{e}^{llm}_{u}), \quad \mathbf{z}_{i}^{llm} = \text{MLP}(\mathbf{e}^{llm}_{i})
\end{equation}
where $\mathbf{e}^{llm}_{u}\in \mathbb{R}^{d_{llm}}$ and $\mathbf{e}^{llm}_{i}\in \mathbb{R}^{d_{llm}}$ represent the text embeddings for user $u$ and item $i$,  $\mathbf{z}_{u}^{llm}\in \mathbb{R}^{d_{rec}}$ and $\mathbf{z}_{i}^{llm}\in \mathbb{R}^{d_{rec}}$ are the projected embeddings ($d_{rec} << d_{llm}$).

To train the projector, we then construct pseudo labels: for each user, items that (1) occupy top-ranked textual embedding similarity scores and (2) have prior interaction with the user are considered reliable labels.
Formally, for each user $u$, we define a set of reliable positive items as follows: 
\begin{equation}
    \mathcal{I}^{al+}_u = \big\{ i | y_{u,i}^{*}=1\big\} \cap \big\{i|\text{Top-}N(\hat{y}^{llm}_{u,i}) \big\},
\end{equation}
where
$\hat{y}^{llm}_{u,i}=\text{sim}(\mathbf{e}^{llm}_{u}, \mathbf{e}^{llm}_{i})$ denote the cosine similarity of LLM-encoded embeddings,
$N$ is a hyperparameter controlling sample quality (typical $N=50$).
The MLP projector is then trained with the following objective,
\begin{scriptsize}
    \begin{equation}\label{eq:embdssl}
        \mathcal{L}_{al} = -\log \frac{\exp(\mathbf{z}_{u} \cdot \mathbf{z}_{i^{al+}} / \tau)}{\exp(\mathbf{z}_{u} \cdot \mathbf{z}_{i^{al+} } / \tau) + \sum_{k=1}^{N} \exp(\mathbf{z}_{u} \cdot \mathbf{z}_{i^{al-}_k} / \tau)},
    \end{equation}    
\end{scriptsize}
where $i^{al+} \in \mathcal{I}^{al+}_u$ and $i^{al-}_k$ are random sampled negatives from $\{ \mathcal{I} \setminus  \mathcal{I}^{al}_u\}$, $\tau=0.5$ is a temperature hyperparameter.
After training the MLP projector, the resulting aligned text embeddings can be formulated as $\mathbf{z'}_{u}^{llm} = \text{MLP}'(\mathbf{e}^{llm}_{u}), \quad \mathbf{z'}_{i}^{llm} = \text{MLP}'(\mathbf{e}^{llm}_{i}),$
where $\mathbf{z'}_{u}^{llm},  \mathbf{z'}_{i}^{llm} \in \mathbb{R}^{d_{rec}}$ and $\text{MLP}'$ denote the traiend projector.

\subsubsection{Semantic-Guided Hard Negative Sampling.}
We leverage the objective-aligned $\mathbf{z'}_{u}^{llm}$ and $\mathbf{z'}_{i}^{llm}$ in negative sampling to select hard negatives and filter out noisy false negatives.
For each $u$, we randomly initialize a hard negative pool $\mathbf{HN}_{u}^{-}=\{j \mid \forall j \in \mathcal{I}_{u}^{-}\}_{k=1}^{K}$ with $K$ negative items.
When training $\text{Rec}_{\theta}$, for each positive $(u,i) \in \mathcal{B}$, we uniformly sample $M$ new negative items by
\begin{equation}
    \mathbf{N}_{u}^{-} = \{j_{m} \mid j_{m} \sim \mathrm{Uniform}(\mathcal{I}_{u}^{-})\}_{m=1}^{M}.
\end{equation}
The $\mathbf{N}_{u}^{-}$ is adopted to update $\mathbf{HN}_{u}^{-}$ dynamically according to the recommender system prediction scores $\hat{y}_{u,i}=\text{Rec}_{\theta}(u,i)$, formally represented as  
\begin{equation}
    \mathbf{HN}_{u}^{-} = \{j_k | j_k \sim P(j) \propto \hat{y}_{u,j}, \forall j \in {\mathbf{HN}_{u}^{-}\cup\mathbf{N}_{u}^{-}}\}_{k=1}^{K},
\end{equation}
where $P(j)$ denotes the sampling distribution.
Considering false negatives might exhibit both high $\hat{y}_{u,j}$ and high semantic similarity, we select the negative item $j$ from $ \mathbf{HN}_{u}^{-}$ with the lowest semantic similarity score,
\begin{equation}
    j = \mathop{\arg\min}\limits_{k \in \mathbf{HN}_{u}^{-}} (s(\mathbf{z'}_{u}^{llm}, \mathbf{z'}_{j_k}^{llm}) ), \forall j_k \in \mathbf{HN}_{u}^{-} 
    \label{eq:negsample}
\end{equation}
where $s(\cdot)$ denotes the cosine similarity.
We then take the hard negative $j$ and positive $(u,i)$ interaction pair to optimize the $\text{Rec}_{\theta}$ with recommendation loss (i.e., BPR loss),
\begin{equation}
    \mathcal{L}_{\text{rec}} = -\frac{1}{|\mathcal{B}|}\sum_{(u,i)\in\mathcal{B}}\log\left(\sigma\left(\hat{y}_{u,i} - \hat{y}_{u,j}\right)\right),
\end{equation}
where the negative item $j$ is selected via Eq.\ref{eq:negsample}.

\subsection{Logical Relevance Guided Interaction Denoising}
With powerful reasoning capability, Large Language Models can infer the logical relationships between users and items that reveal users' potential interest in items.
Therefore, we design the following strategies that take into account these LLM-inferred logical relevance within user-item interactions to identify noisy and hard samples.

\subsubsection{Logical Relevance Inference.}
We first obtain the logical relevance between users and items in RS with LLM.
Given the enormous number of $u$ and $i$, employing LLMs to scrutinize every user-item pair is infeasible.
Therefore, we select potential hard and noisy interactions before subjecting them to LLM for logical relevance analysis.
Specifically, we leverage a trained recommender system $\text{Rec}_{pre}$ to sample user-item pairs.
For each user $u$, two subsets are constructed:
(1) High-Score Negatives: $n_{1}$ items that have higher prediction scores $\hat{y}^{pre}_{u,i}= \text{Rec}_{pre}(u,i)$, where $y_{u,i}^{*}=0$.
(2) Low-Score Positives: $n_{2}$ items that have lower prediction scores $\hat{y}^{pre}_{u,i}$ and $y_{u,i}^{*}=1$.
The unified candidate set $\mathcal{C}$ is defined as:
\begin{scriptsize}
\begin{equation}
\mathcal{C} = \underset{u \in \mathcal{U}}{\cup} (\underbrace{\{(u, i)|i \sim P_{\mathcal{I}_u^-}(i) \propto \hat{y}_{u,i}^{pre}\}}_{\text{High-score false negatives}} \cup \underbrace{\{(u, i) | i \sim P_{\mathcal{I}_u^+}(i) \propto - \hat{y}_{u,i}^{pre}\}}_{\text{Low-score false positives}}),
\end{equation}
\end{scriptsize}
where $\mathcal{I}_u^{-} = \{ i \mid y_{u,i}^* = 0 \}$, $\mathcal{I}_u^{+} = \{ i \mid y_{u,i}^* = 1 \}$, $P_{\mathcal{S}}(i)$ denotes the sampling distribution on the set $\mathcal{S}$.

For each candidate user-item pair $(u, i) \in C$, we assess their relevance from two aspects:
(1) User-based Rating: This employs the user text profile to describe user $u$'s preference.
LLM is prompted with a predefined prompt template $\mathcal{T}^{\text{user}}$ to rate the logical relevance between $u$ and $i$.
\begin{equation}
    r_{u,i}^{\text{user}} = \text{LLM}( \mathcal{T}^{\text{user}}(V_{u}^{\text{user}}, \mathcal{P}_i)),
\end{equation}
where $r_{u,i}^{\text{user}} \in [\text{High, Mid, Low}]$ and $V_{u}^{\text{user}}=\mathcal{P}_{u}$.
(2) Item-based Rating: This takes the profiles of item that $u$ has interacted and has high $\hat{y}^{pre}_{u,i}$ as $u$'s preference descriptions.
Using predefined prompt templates $\mathcal{T}^{\text{item}}$, LLM rates the logical relevance between $u$ and $i$ as follows,
\begin{equation}
    r_{u,i}^{\text{item}} = \text{LLM}( \mathcal{T}^{\text{item}}(V_{u}^{\text{item}} , \mathcal{P}_i) ),
\end{equation}
where $r_{u,i}^{\text{item}} \in [\text{High, Mid, Low}]$, $V_{u}^{\text{item}} = \{ \mathcal{P}_j | j \in \mathcal{I},  y^{*}_{u,j} = 1, \hat{y}^{pre}_{u,j} \in \text{top-K}(\hat{\mathbf{y}}^{pre}_{u})\}$, $\hat{\mathbf{y}}^{pre}_{u}$ denotes the preference scores of $u$ with all items predicted by $\text{Rec}_{pre}$.

\subsubsection{Interaction Denoising.}
Building on the logical relevance rates $r_{u,i}^{\text{user}}$ and $r_{u,i}^{\text{item}}$, we identify hard and noisy samples within the candidate set $\mathcal{C}$.
To preserve performance-enhancing hard samples while conservatively filtering noise, we define the noise subset $\mathcal{C}_{\text{N}}$ and hard subset $\mathcal{C}_{\text{H}}$ as follows,
\begin{equation}
    \mathcal{C}_{\text{H}} = \{(u,i) \in \mathcal{C} \mid r_{u,i}^{\text{user}} = \text{High} \land r_{u,i}^{\text{item}} = \text{High}\}
\end{equation}
\begin{equation}
    \mathcal{C}_{\text{N}} = \mathcal{C} \setminus \mathcal{C}_{\text{H}}
\end{equation}
That is, $(u,i)$ interactions are considered as hard samples if both the User-Centric and Item-Centric ratings yield High logical relevance scores.
All remaining samples in $\mathcal{C}$ are treated as noisy samples.
As integrate $\mathcal{C}_{\text{H}}$ and $\mathcal{C}_{\text{N}}$ to train recommenders might introduce label noise.
We construct the user-item interaction graph $G'=\{\mathcal{U}, \mathcal{I}, \mathcal{E}'\}$ with the original interaction graph $G=\{\mathcal{U}, \mathcal{I}, \mathcal{E}\}$ to guide interaction denoise.
Here, the edge $\mathcal{E}'$ can be formally formulated as 
\begin{equation}
    \mathcal{E}' = \mathcal{E} \setminus \{e_{u,i} \mid (u,i) \in \mathcal{C}_{\text{N}}\} \cup \{e_{u,i} \mid (u,i) \in \mathcal{C}_{\text{H}}\},
\end{equation}
where $e_{u,i}$ denotes an edge between user $u$ and item $i$.
We obtain user ($\mathbf{z}_{u}, \mathbf{z}'_{u}$) and item ($\mathbf{z}_{i}, \mathbf{z}'_{i}$) representations from both $G$ and $G'$ with the recommender $\text{Rec}_{\theta}$, all in $\mathbb{R}^{d_{rec}}$, 
\begin{align}
    \mathbf{z}_{u}, \mathbf{z}_{i} = \text{Rec}_{\theta}(G); \quad
    \mathbf{z}'_{u}, \mathbf{z}'_{i} = \text{Rec}_{\theta}(G').
\end{align}
A cross-graph contrastive alignment strategy is designed to enhance $(u,i)$ interactions that are consistent on $G$ and $G'$,
\begin{equation}
    \mathcal{L}_{\text{de}}^{u'i} = -\frac{1}{|\mathcal{B}|} \sum_{(u,i)\in\mathcal{B}} \log \frac{\exp(s(\mathbf{z'}_{u}, \mathbf{z}_{i})/\tau_{\text{de}})}{\sum\limits_{(u,j)\in\mathcal{B}} \exp(s(\mathbf{z}'_{u},\mathbf{z}_{j})/\tau_{\text{de}})}
\end{equation}
where $s(\cdot,\cdot)$ is cosine similarity, and $\tau_{de}$ is temperature hyperparameter.
Together with the item side loss, the denoise loss is $\mathcal{L}_{\text{de}} = \mathcal{L}_{\text{de}}^{u'i}+\mathcal{L}_{\text{de}}^{ui'}.$
As all positive $(u,i)$ pairs ($i\in\mathcal{I}_{u}^{+}$) might appear in both the numerator and the denominator of $\mathcal{L}_{\text{de}}$,
the embedding of $(u,i)$ pairs that consistent with $G$ and $G'$ (i.e., high similarity between $(\mathbf{z}_{u}, \mathbf{z}'_{i})$ and $(\mathbf{z}'_{u}, \mathbf{z}_{i})$)  will be aligned better, while those inconsistent are suppressed.

\subsubsection{Hallucination-Robust Contrastive Learning.}
Although $G'$ are constructed based on LLM-inferred interaction, leveraging $\mathcal{C}_{\text{H}}$ and $\mathcal{C}_{\text{N}}$ risks propagate hallucination-induced interactions.
Therefore, we design a graph contrastive learning strategy to reduce the negative impact of hallucination-induced edges in $G'$.
In each training batch, we generate two augmented views by stochastic edge drop to $G'$ and $G$:
\begin{equation}
    G_{\mathrm{aug}} = \left(\mathcal{U}, \mathcal{I}, \mathcal{E} \backslash \mathcal{M}\right);\quad
    G'_{\mathrm{aug}} = \left(\mathcal{U}, \mathcal{I}, \mathcal{E}^{\prime} \backslash \mathcal{M'}\right),
\end{equation}
where $\mathcal{M}',\mathcal{M} \sim \operatorname{Bernoulli}(\rho; |\mathcal{E}'|)$ denotes the set of randomly masked edges.
Each edge is dropped independently with probability $\rho$.
Both $G'_{\mathrm{aug}}$ and $G_{\mathrm{aug}}$ are processed by the same graph-based recommendation models $\text{Rec}_{\theta}$ with parameters $\theta$, generating user and item representations:
\begin{align}
    \mathbf{z}_{u}^{(1)}, \mathbf{z}_{i}^{(1)} = \text{Rec}_{\theta}(G'_{\mathrm{aug}});
    \mathbf{z}_{u}^{(2)}, \mathbf{z}_{i}^{(2)} = \text{Rec}_{\theta}(G_{\mathrm{aug}}).
\end{align}
where $\mathbf{z}_{u}^{(k)}, \mathbf{z}_{i}^{(k)}$ are the user and item representations ($k=[1,2]$).
We adopt contrastive loss to maximize the agreement of positive pairs and minimize that of negative pairs,
\begin{equation}
    \mathcal{L}_{\text{hal}}^{\text{user}} = \sum_{u \in \mathcal{U}} -\log \frac{\exp(s(\mathbf{z}_{u}^{(1)}, \mathbf{z}_{u}^{(2)})/\tau_{\text{hal}})}{\sum_{v \in \mathcal{U}} \exp(s(\mathbf{z}_{u}^{(1)}, \mathbf{z}_{v}^{(2)})/\tau_{\text{hal}})}, 
\end{equation}
where $s(\cdot)$ denotes the cosine similarity; $\tau_{\text{hal}}$ is the temperature hyperparameter.
We get the objective function by combining the item side loss $\mathcal{L}_{\text{hal}} = \mathcal{L}_{\text{hal}}^{\text{user}} + \mathcal{L}_{\text{hal}}^{\text{item}}$.

\subsubsection{Joint Optimization.}
We optimize the recommender system model with the total loss:
\begin{equation}
\mathcal{L}_{\text{total}} = \mathcal{L}_{\text{rec}} + \lambda_1 \mathcal{L}_{\text{de}} + \lambda_2 \mathcal{L}_{\text{hal}},
\end{equation}
where $\lambda_1, \lambda_2$ are hyperparameters that balance the weight.

\section{Experiment}
\begin{table*}[!t]
    \centering
    \setlength{\tabcolsep}{3.1pt}
    \begin{tabularx}{\textwidth}{c|c|cccc|cccc|cccc}
        \toprule
        \multicolumn{2}{c|}{\textbf{Dataset}} & \multicolumn{4}{c|}{\textbf{Amazon-book}} & \multicolumn{4}{c|}{\textbf{Yelp}} & \multicolumn{4}{c}{\textbf{Steam}}\\
        \midrule
        \textbf{Backbone} & \textbf{Method} & \textbf{R@10} & \textbf{N@10} & \textbf{R@20} & \textbf{N@20} & \textbf{R@10} & \textbf{N@10} & \textbf{R@20} & \textbf{N@20} & \textbf{R@10} & \textbf{N@10} & \textbf{R@20} & \textbf{N@20} \\
        \midrule\midrule
        
        \multirow{10}*{NGCF} & Normal & 0.0763 & 0.0584 & 0.1204 & 0.0726 & 0.0634 & 0.0527 & 0.1045 & 0.0664 & 0.0795 & 0.0658 & 0.1271 & 0.0814  \\
        ~ & WBPR & 0.0765 & 0.0587 & 0.1212 & 0.0729 & 0.0636 & 0.0530 & 0.1048 & 0.0669 & 0.0796 & 0.0657 & 0.1270 & 0.0813  \\
        ~ & T-CE & 0.0844 & 0.0648 & 0.1288 & 0.0789 & 0.0650 & 0.0543 & 0.1071 & 0.0683 & 0.0808 & 0.0671 & 0.1290 & 0.0828 \\
        ~ & BOD & 0.1199 & 0.0959 & 0.1666 & 0.1109 & 0.0721 & 0.0615 & 0.1193 & 0.0770 & 0.0921 & 0.0767 & 0.1461 & 0.0944  \\
        ~ & SGL & 0.0902 & 0.0707 & 0.1362 & 0.0856 & 0.0667 & 0.0557 & 0.1103 & 0.0704 & 0.0832 & 0.0689 & 0.1305 & 0.0843  \\
        ~ & SimGCL & 0.0863 & 0.0667 & 0.1304 & 0.0807 & 0.0680 & 0.0577 & 0.1140 & 0.0732 & 0.0819 & 0.0680 & 0.1274 & 0.0830  \\
        ~ & XSimGCL & 0.0963 & 0.0746 & 0.1431 & 0.0894 & 0.0701 & 0.0598 & 0.1153 & 0.0744 & 0.0957 & 0.0710 & 0.1352 & 0.0871  \\
        ~ & RLMRec & 0.0855 & 0.0635 & 0.1323 & 0.0815 & 0.0697 & 0.0568 & 0.1150 & 0.0723 & 0.0865 & 0.0690 & 0.1370 & 0.0862  \\
        ~ & LLaRD & \underline{0.1265} & \underline{0.1021} & \underline{0.1834} & \underline{0.1203} & \underline{0.0845} & \underline{0.0758} & \underline{0.1347} & \underline{0.0903} & \underline{0.0997} & \underline{0.0821} & \underline{0.1565} & \underline{0.1012}  \\
        \cmidrule{2-14}
        \rule{0pt}{9pt} ~ & \textbf{$\text{LLMHNI}$} & \textbf{0.1290} & \textbf{0.1038} & \textbf{0.1852} & \textbf{0.1230} & \textbf{0.0880} & \textbf{0.0823} & \textbf{0.1367} & \textbf{0.0938} & \textbf{0.1037} & \textbf{0.0864} & \textbf{0.1588} & \textbf{0.1057}  \\
        
        \midrule\midrule
        \multirow{10}*{LightGCN} & Normal & 0.0950 & 0.0746 & 0.1415 & 0.0893 & 0.0617 & 0.0529 & 0.1011 & 0.0659 & 0.0838 & 0.0701 & 0.1317 & 0.0858 \\
        ~ & WBPR & 0.0956 & 0.0749 & 0.1419 & 0.0901 & 0.0618 & 0.0531 & 0.1015 & 0.0662 & 0.0840 & 0.0704 & 0.1312 & 0.0850  \\
        ~ & T-CE & 0.0990 & 0.0779 & 0.1499 & 0.0939 & 0.0740 & 0.0623 & 0.1216 & 0.0779 & 0.0877 & 0.0731 & 0.1376 & 0.0893  \\
        ~ & BOD & 0.1273 & 0.0996 & 0.1792 & 0.1162 & 0.0843 & 0.0714 & 0.1355 & 0.0882 & 0.0955 & 0.0791 & 0.1484 & 0.0966  \\
        ~ & SGL & 0.1091 & 0.0872 & 0.1588 & 0.1030 & 0.0762 & 0.0648 & 0.1235 & 0.0803 & 0.0890 & 0.0743 & 0.1378 & 0.0903  \\
        ~ & SimGCL & 0.1172 & 0.0940 & 0.1681 & 0.1102 & 0.0785 & 0.0669 & 0.1265 & 0.0827 & 0.0887 & 0.0738 & 0.1373 & 0.0899  \\
        ~ & XSimGCL & 0.1153 & 0.0931 & 0.1637 & 0.1084 & 0.0769 & 0.0663 & 0.1277 & 0.0830 & 0.0884 & 0.0736 & 0.1385 & 0.0903  \\
        ~ & RLMRec & 0.1034 & 0.0788 & 0.1601 & 0.0960 & 0.0794 & 0.0652 & 0.1275 & 0.0815 & 0.0926 & 0.0746 & 0.1452 & 0.0924 \\
        ~ & LLaRD & \underline{0.1408} & \underline{0.1126} & \underline{0.2028} & \underline{0.1326} & \underline{0.0975} & \underline{0.0809} & \underline{0.1574} & \underline{0.1008} & \underline{0.1054} & \underline{0.0868} & \underline{0.1631} & \underline{0.1059}  \\
        \cmidrule{2-14}
        \rule{0pt}{9pt} ~ & \textbf{$\text{LLMHNI}$} & \textbf{0.1423} & \textbf{0.1168} & \textbf{0.2040} & \textbf{0.1369} & \textbf{0.0981} & \textbf{0.0837} & \textbf{0.1594} & \textbf{0.1047} & \textbf{0.1065} & \textbf{0.0893} & \textbf{0.1646} & \textbf{0.1087} \\
        \bottomrule
        \end{tabularx}
    \caption{
        Performance comparison of backbone recommenders trained with different denoising approaches.
        R and N refer to Recall and NDCG, respectively.
        The highest scores are in \textbf{bold}, and the runner-ups are with \underline{underline}.
        All results are statistically significant according to the t-tests with a significance level of $p < 0.01$.
    }
   \label{table:performance}
\end{table*}

We compare LLMHNI with state-of-the-art denoise methods on two backbones and three real-world datasets.
Experiments are directed by following research questions (RQs):
\begin{itemize}
    \item \textbf{RQ1:} How does LLMHNI performs compared with other state-of-the-art denoise methods across datasets?
    \item \textbf{RQ2:} Does the LLMHNI demonstrate robustness when tackling different levels of noisy data?
    \item \textbf{RQ3:} What is the effect of different components within the LLMHNI on performance?
    \item \textbf{RQ4:} How do hyperparameters in LLMHNI influence the effectiveness?
    \item \textbf{RQ5:} What is the training efficiency of LLMHNI?
\end{itemize}

\subsection{Experiment Settings}

\subsubsection{Datasets.}
We conduct all experiments on three datasets:
(1) \textbf{Amazon-Books} collected from the Amazon platform. We conduct experiments on the book subcategories.
(2) \textbf{Yelp} is a large-scale dataset with real check-in history.
(3) \textbf{Steam} consists of users and games on the Steam platform.
Since we adopt the item and user profile provided in \cite{Ren2023RepresentationLW}, we process these datasets following their settings.

\subsubsection{Evaluation Metrics.}
Following existing works on recommender system denoising \cite{Wang2021ImplicitFA, he2024double}, we report the results w.r.t. two widely used metrics: NDCG@\textit{K} and Recall@\textit{K} (\textit{K}$=[10,20]$).

\subsubsection{Baselines.}
We conduct experiments with the NGCF \cite{Wang2019NeuralGC} and LightGCN \cite{He2020LightGCN} backbones.
Three types of denoising approaches are compared:
(1) Instance-level approaches, including WBPR \cite{Gantner2012PersonalizedRF}, T-CE \cite{Wang2021DenoisingIF} and BOD \cite{Wang2022EfficientBO}.  
(2) Representation-level approaches, including SGL \cite{Wu2020SelfsupervisedGL}, SimGCL \cite{yu2022graph} and XSimGCL \cite{yu2023xsimgcl}.
(3) LLM Enhanced approaches, including RLMRec \cite{Ren2023RepresentationLW} and LLaRD \cite{wang2025unleashing}.

\subsubsection{Implementation Details.}
For all models, the embedding size is set to 64, the batch size is 1024, and the learning rate is 1e-3.
All models are trained with the Adam optimizer.
For baseline models, we refer to their best parameter setups reported in original papers.
For our model, we set “gpt-4o-2024-08-06” as $\text{LLM}$ and “text-embedding-ada-002” as $\text{LLM}_{enc}$.
We set the uniform sampled negative item number $M$ at 30 and the hard negative candidate number $K$ at 10.

\subsection{Performance Comparison (RQ1)}
To evaluate the effectiveness and generalizability of our propsoed framework, we compared our LLMHNI with existing denoising baselines across three datasets and two backbone models.
The result is shown in Table \ref{table:performance}.
Our LLMHNI consistently exceeds desnoising baselines in all three datasets and both backbone models.
On average, LLMHNI achieves 46.55\% improvements on the vanilla NGCF backbone and 45.31\% on the original LightGCN backbone.
Compared with previous instance-level denoising approaches (i.e., T-CE and BOD) and representation-level techniques (i.e., SGL, SimGCL, and XSimGCL), LLMHNI exhibits a substantial performance improvement from 11.78\% to 37.73\%.
This significant enhancement is attributed to our utilization of LLMs to provide auxiliary relevance signals beyond the original interaction data.
Regarding LLM-enhanced denoising techniques (i.e., RLMRec and LLaRD), LLM outperforms them by roughly 2.47\% to 33.86\%.
Although RLMRec and LLaRD incorporate supplementary information generated by LLMs, they lack the capabilities of identifying hard samples.
Our LLMHNI, in comparison, extends the LLM-provided signals in hard sample identification, thereby excelling both baselines.

\subsection{Noise Robustness (RQ2)}

\begin{figure}[t]
    \centering
    \includegraphics[width=\linewidth]{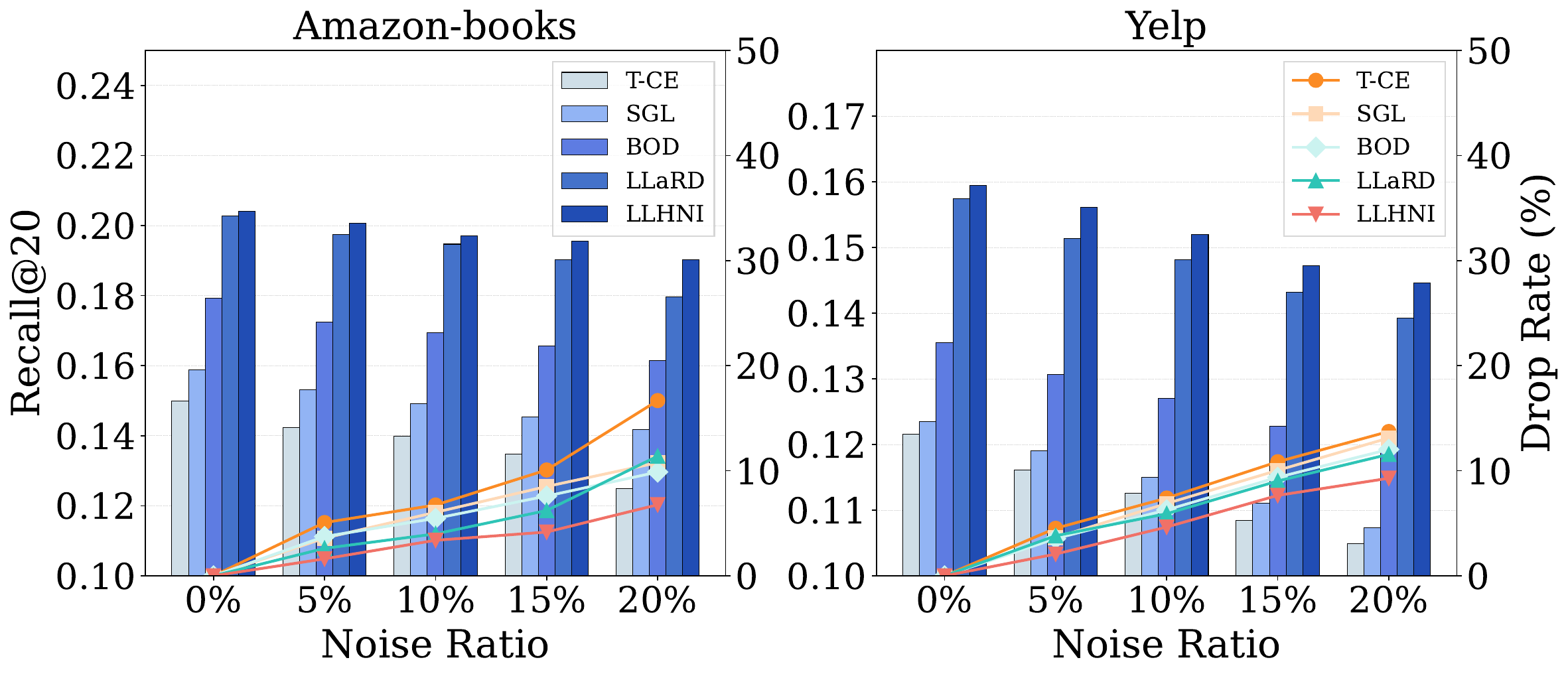}
    \caption{Model performance $w.r.t$ different noise ratio. The bar chart represents Recall values (see left y-axis), while the line chart shows Drop Rate (see right y-axis) All denoise methods are trained with the LightGCN backbone.}
    \label{fig:Noiserobust}
\end{figure}

To assess the robustness of LLMHNI's noise resistance capabilities, following previous methods \cite{Ren2023RepresentationLW, Wang2022EfficientBO}, we add certain levels of non-existent interactions to the training set (i.e., 5\%, 10\%, 15\%, 20\% negative interactions) and keep the test set unchanged.
Fig. \ref{fig:Noiserobust} shows the results in the Amazon-books and Yelp.
Our LLMHNI consistently outperforms other baseline models across all noise levels.
While performance drops as noise levels rise, the rate at which LLMHNI's performance declines remains relatively stable compared to other baselines, demonstrating that LLMHNI is the least impacted by noise.
This indicates that LLMHNI effectively identifies noisy and hard samples, even in the presence of significant noise.

\subsection{In-depth Analysis of LLMHNI (RQ3 - RQ5)}

\subsubsection{Ablation Study (RQ3).}
To assess the impact of each component within LLMHNI, we conducted ablation studies with four variants.
Here, SR represents components associated with semantic relevance, and LR pertains to logical relevance:
\textbf{(1) w/o $\text{SR}_{\text{lmns}}$}: Replaces LLM-embedding guided hard negative sampling (Equation \ref{eq:negsample}) with uniform sampling.
\textbf{(2) w/o $\text{SR}_{\text{al}}$}: Excludes objective alignment strategy applied to LLM-encoded embeddings (Equation \ref{eq:embdssl}).
\textbf{(3) w/o $\text{LR}_{\text{hal}}$}: Removes the graph contrastive loss $\mathcal{L}_{\text{hal}}$ aimed at mitigating unreliable interactions.
\textbf{(4) w/o $\text{LR}_{\text{de}}$}: Removes the graph contrastive loss $\mathcal{L}_{\text{de}}$ for cross-graph user-item alignment.
Table \ref{tab:ablation} shows varying performance degradation when specific modules are removed. 
The drop in performance for \textbf{w/o $\text{TR}_{\text{lmns}}$} underscores the crucial role of auxiliary semantic relevance in distinguishing hard and false negatives.
Similarly, the reduction in performance for \textbf{w/o $\text{TR}_{\text{al}}$} illustrates the significance of objective-aligned LLM-encoded embeddings in selecting hard negative items.
Furthermore, performance declines in \textbf{w/o $\text{LR}_{\text{de}}$} highlight the effectiveness of logical relevance in interaction denoising.
While the performance drop in \textbf{w/o $\text{LR}_{\text{hal}}$} demonstrates the importance of eliminating hallucination-induced interaction graph edges.

\begin{table}[t]
    \setlength{\tabcolsep}{2.5pt}
    \small
    \begin{tabular*}{\linewidth}{l|cccc|c}
        \toprule
        Variants & w/o TR$_{\text{lmns}}$  & w/o TR$_{\text{ssl}}$ & w/o LR$_{\text{nie}}$ & w/o LR$_{\text{uis}}$ & \textbf{LLMHNI} \\
        \midrule
        R@10 & 0.1199 & 0.1248  & 0.1125 & 0.1294 & 0.1423 \\
        R@20 & 0.1799 & 0.1848  & 0.1772 & 0.1854 & 0.2040 \\
        N@10 & 0.0937 & 0.0982  & 0.0855 & 0.1047 & 0.1168 \\
        N@20 & 0.1112 & 0.1174  & 0.1060 & 0.1230 & 0.1369 \\
        \bottomrule
    \end{tabular*}
    \caption{The effect of components in LLMHNI with the LightGCN on Amazon-books datasets.}
    \label{tab:ablation}
\end{table}

\subsubsection{Hyperparameters Analysis (RQ4).}

\begin{figure}[t]
    \centering
    \includegraphics[width=0.9\linewidth]{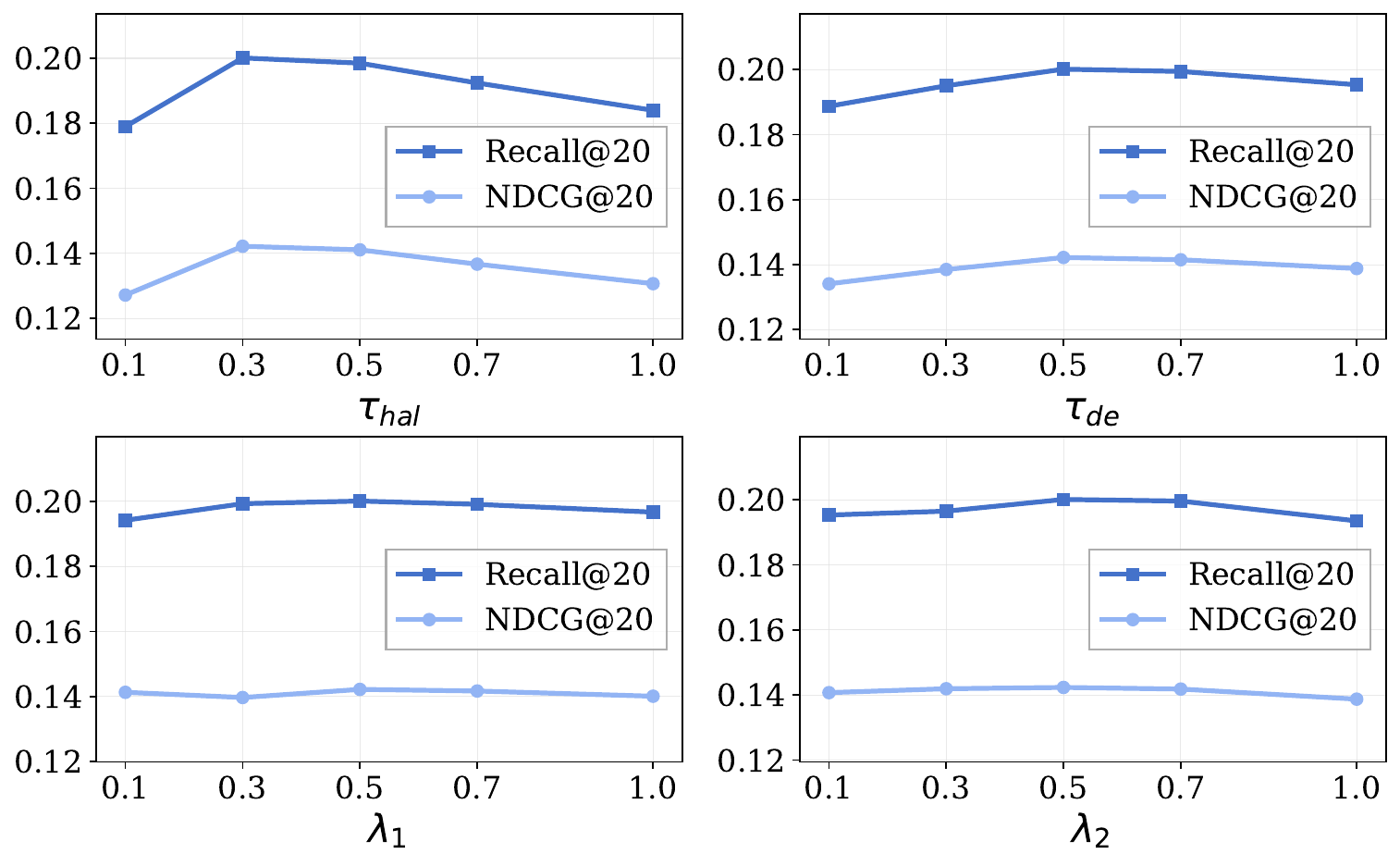}
    \caption{Hyper-parameter analysis of $\text{LLMHNI}$ with LightGCN backbone on the Amazon-books datasets.}
    \label{fig:Hyperparams}
\end{figure}

To assess LLMHNI's sensitivity to hyperparameter changes, we varied the hyperparameters $\lambda_{1}$, $\lambda_{2}$, $\tau_{de}$, and $\tau_{hal}$ within the range of $[0.1, 0.3, 0.5, 0.7, 1.0]$.
Due to space constraints, we present only the results from the Amazon-books dataset in Figure \ref{fig:Hyperparams} as results from other datasets show similar patterns.
Our analysis reveals that modifications in temperature parameters $\tau_{hal}$ and $\tau_{de}$ lead to significant performance variations.
The performance of both $\tau_{hal}$ and $\tau_{de}$ shows an upward trend first and then drops steadily.
This highlights the importance of choosing a suitable temperature in contrastive learning.
Furthermore, alterations in $\lambda_{1}$ and $\lambda_{2}$ have minimal impact on performance, demonstrating the stability of LLMHNI with these two hyperparameters.

\subsubsection{Training Efficiency Analysis (RQ5).}
While LLMHNI includes multiple components, its overall time complexity remains comparable to mainstream denoising methods.
The relevance signal generation is conducted before training the recommender system; thus is excluded from this analysis.
The calculation of $\mathcal{L}_{\text{hal}}$ introduce an additional complexity of $O(|\mathcal{E}|d_{rec}(2+|\mathcal{V}|))$, where $\mathcal{V}= \mathcal{U} \cup \mathcal{I}$.
While $\mathcal{L}_{\text{de}}$ takes in-batch negatives, resulting in $O\left(|\mathcal{E}|d_{rec}(2+2\mathcal{B})\right)$ time cost.
In addition to the inference on interaction graph $G$, our method introduces additional inference on $G'_{aug}$ and $G_{aug}$, resulting in addition complexity of $O(|\mathcal{E'}_{aug}| d_{rec} L)$, and $O(|\mathcal{E}_{aug}| d_{rec} L)$, respectively.
The $L$ denotes the number of layers in the backbone recommenders.
We also provide a running time comparison with the baselines on the LightGCN backbone.
Table \ref{tab:time} presents the results obtained on a server with two Intel(R) Xeon(R) Gold 5118 CPUs (12 cores each) and an NVIDIA GeForce RTX 3090 GPU.

\begin{table}[t]
    \small
    \centering
    \begin{tabular*}{0.82\linewidth}{l|ccc}
        \toprule
        \textbf{Method} & \textbf{Amazon-Book} & \textbf{Yelp} & \textbf{Steam }\\
        \midrule
        Normal & 1.0307 & 1.3568 & 4.7756 \\
        \midrule
        T-CE & 1.2369 & 1.4639 & 5.0580 \\
        XSimGCL & 1.5066 & 1.8678 & 7.0733 \\
        SimGCL & 3.0946 & 4.0361 & 13.7971 \\
        SGL & 3.5637 & 4.4746 & 14.7370 \\
        BOD & 5.7516 & 6.4349 & 19.5580 \\
        \midrule
        LLMHNI & 6.9677 & 8.7488 & 26.6929 \\
        \bottomrule
    \end{tabular*}
    \caption{Comparison of training time in seconds per epoch across different datasets and baseline denoise methods.}
    \label{tab:time}
\end{table}

\section{Related Works}
\subsection{Denoise Recommendation}
Recommenders are pointed out to be affected by users’ unconscious behaviors ~\cite{Wang2021ClicksCB}, leading to noisy data.
As a result, many efforts are designed to alleviate the problem.
These approaches can be categorized into two paradigms: sample dropping ~\cite{Gantner2012PersonalizedRF,Lin2023AutoDenoiseAD} and sample re-weighting ~\cite{Wang2022EfficientBO, Gao2022SelfGuidedLT}.
Sample dropping methods aim to keep clean samples and discard noisy ones.
For instance, T-CE ~\cite{Wang2021DenoisingIF} observes that noisy samples exhibit high loss values and removes them during training.
IR~\cite{Wang2021ImplicitFA} iteratively generates pseudo-labels to discover noisy examples.
DCF~\cite{he2024double} introduces a double correction method that drops samples based on loss values and prediction score variances.
Sample re-weighting methods try to mitigate the impact of noisy samples by assigning lower weights to them.
Typically,
R-CE ~\cite{Wang2021DenoisingIF} assigns lower weights to noisy samples according to the prediction score.
BOD ~\cite{Wang2022EfficientBO} considers weight assignment as a bi-level optimization problem.
Despite their promising results, they rely on data patterns to recognize noisy samples (e.g., loss values and prediction scores),
resulting in the hard-noisy sample confusion issue.

\subsection{LLMs for Recommendation}
LLMs are effective tools for NLP tasks and have gained significant attention in the domain of Recommendation Systems (RS).
For the adaption of LLMs in recommendations, existing works can be divided into three categories ~\cite{Wu2023ASO}: LLM as RS, LLM Embedding for RS, and LLM token for RS.
The LLM as RS aims to transform LLMs into effective recommendation systems ~\cite{chao2024makelargelanguagemodel}, such as LC-Rec ~\cite{Zheng2023AdaptingLL} and LLM-TRSR ~\cite{Zheng2024HarnessingLL}.
In contrast, the LLM embedding for RS and LLM token for RS views the language model as an enhancer.
The former typically adopts embeddings related to users and items, incorporating semantic information in the recommender ~\cite{Ren2023RepresentationLW}.
While the latter generates text tokens to capture potential preferences between user and items ~\cite{Wei2023LLMRec,Xi2023TowardsOR}.
Recent studies also leverage LLMs in recommender system denoise, where RLMRec \cite{Ren2023RepresentationLW} and DALR \cite{peng2025denoising} implicitly eliminate noise at the representation-level.
The LLaRD \cite{wang2025unleashing} takes LLMs to generate preference knowledge and relationship knowledge to denoise.
However, none of them discuss the potential of LLMs in supporting the identification of hard and noisy samples.

\section{Conclusion}

In this work, we investigate the potential of Large Language Models in solving the hard and noisy sample confusion in recommender systems.
We discvoered that LLMs can offer valuable auxiliary signals for addressing hard-noisy sample confusion, including the user-item semantic relevance from LLM-encoded embeddings and the user-item logical relevance from LLM-inferred interactions.
To take advantage of these two signals, we introduce the \textbf{L}arge \textbf{L}anguage \textbf{M}odel Enhanced \textbf{H}ard-\textbf{N}oisy Sample \textbf{I}dentification framework (\textbf{LLMHNI}).
LLMHNI generates both relevance signals, leveraging them to resolving hard-noisy confusion issues in both hard negative sampling and interaction denosing.
More importantly, LLMHNI enhances the utilization of these two signals in recommender systems by effectively addressing the objective mismatch of LLM-encoded embeddings and hallucinations in LLM-inferred interactions.
Experiments on three real-world datasets and two backbone recommenders confirm the efficacy of our approach.

\section{Acknowledgments}
This work was supported by the National Natural Science Foundation of China (Grant No. 62572417, No.92370204), National Key R\&D Program of China (Grant No.2023YFF0725004), the Guangzhou Basic and Applied Basic Research Program under Grant No. 2024A04J3279, and Education Bureau of Guangzhou Municipality.

\bibliography{aaai2026}

\clearpage
\section{Supplementary Materials}

\subsection{Details of Easy, Hard and Noisy Samples in Fig. \ref{fig:Introduction}}
Here we provide details of how we construct and obtain the \textit{\textbf{Easy}}, \textit{\textbf{Hard}}, and \textit{\textbf{Noisy}} loss values and prediction scores.
For the \textit{\textbf{loss value scores}}, we obtain the \textit{\textbf{Easy}} sample values by sampling negative items uniformly and calculating the corresponding loss values.
In contrast, those \textit{\textbf{Hard}} samples are selected by sampling $n$ ($n=3$) negative items for each $(u,i)$ pair and select the one with the highest prediction scores (i.e., hard negative) as the final negative when calculating training loss.
The \textit{\textbf{Noisy}} sample loss values are collected by taking positive items in the test set as the negative item when calculating the BPR loss, thereby the loss value can be considered as negative.
For the \textit{\textbf{prediction scores}}, the \textit{\textbf{Easy}} sample set and \textbf{\textit{Hard}} sample set are constructed based on the Katz Index as follows,
\begin{equation}
    Katz(u,i)=\sum_{l=1}^{\infty}\beta^{l}\left|\operatorname{paths}_{x, y}^{l}\right|=\sum_{l=1}^{\infty}\beta^{l}(A^{l})_{x, y},
\end{equation}
where the maximum path length $l$ is set to 3 and $\beta = 0.5$ is the weight decay value. 
We select $(u,i)$ pairs with higher Katz Index as easy samples while those with lower Katz index are hard samples.
The prediction score of \textbf{\textit{Noisy}} samples are collected by calculating the prediction scores of those non-existence user-item interactions.
In this way, we obtain both the loss values and the prediction scores of \textbf{\textit{Easy}}, \textbf{\textit{Hard}} and \textbf{\textit{Noisy}} samples.

\subsection{Additional Experiment Detials}

\subsubsection{Datasets Statistics.}
In this section, we provide details about the preprocessed dataset used in the experiment. We utilize the three datasets from RLMRec \cite{Ren2023RepresentationLW}, where each item and user includes a corresponding text profile. We adhere to their preprocessing methods, specifically filtering out interactions with ratings below 3 in both the Amazon-books and Yelp datasets, while no rating-based filtering is applied to the Steam dataset. Additionally, k-core filtering is performed, and the data is split into training, validation, and test sets in a 3:1:1 ratio. The statistics of the preprocessed datasets are presented in Table \ref{tab:dataset}.

\begin{table}[h]
    \centering
    \setlength{\tabcolsep}{3.5pt}
    \small
    \begin{tabular*}{\linewidth}{c|cccc}
        \toprule
        Datasets & \# Users & \# Items & \# Interactions & \# Sparsity \\
        \midrule
        Amazon-books & 11,000 & 9,332 & 120,464 & 99.88\% \\
        Yelp & 11,091  & 11,010 & 166,620 & 99.86\% \\
        Steam & 23,310  & 5,237 & 316,190 & 99.74\% \\
        \bottomrule
    \end{tabular*}
    \caption{Statistics of preprocessed datasets.}
    \label{tab:dataset}
\end{table}

\subsubsection{Baselines and Backbone Models}
We provide details about the backbone models and denoise baselines we utilized in the experiment.
Our experiment are conducted with two backbone models
\begin{itemize}
    \item \textbf{NGCF} \shortcite{Wang2019NeuralGC} models the user-item interaction graph with GNN for collaborative filtering.
    \item \textbf{LightGCN} \shortcite{He2020LightGCN} is a widely adopted graph-based recommendation model that removes the feature transformation and non-linear activation in NGCF, achieving better effectiveness and efficiency.
\end{itemize}
Our baseline denoise methods include instance-level denoising methods, representation-level denoising methods, and the LLM-enhanced denoising methods.
(1) Instance-level Denoising, including 
\begin{itemize}
    \item \textbf{WBPR} \shortcite{Gantner2012PersonalizedRF} is a sampling-based denoising method that assumes a not-interacted but highly popular item should be assigned higher weights in negative sampling.
    \item \textbf{T-CE} \shortcite{Wang2021DenoisingIF} is a sample-dropping method that removes the samples with higher loss by a dynamic threshold.
    \item \textbf{BOD} \shortcite{Wang2022EfficientBO} models denoising as a bi-level optimization problem and optimize a generator to assign weights for training samples.
\end{itemize}
(2) Representation-level denoising, including 
\begin{itemize}
    \item \textbf{SGL} \shortcite{Wu2020SelfsupervisedGL} is a self-supervised graph contrastive learning framework with multiple views for robust representations.
    \item \textbf{SimGCL} \shortcite{yu2022graph} is a graph contrastive learning framework that adds random noise to embeddings to create contrastive views.
    \item \textbf{XSimGCL} \shortcite{yu2023xsimgcl} optimize the SimGCL for more efficient graph contrastive learning in recommendation.
\end{itemize}
(3) LLM Enhanced denoising approaches, including
\begin{itemize}
    \item \textbf{RLMRec} \shortcite{Ren2023RepresentationLW} utilizes LLMs to encode the complex user behavior semantics, enhancing recommendations through contrastive and generative techniques.
    \item \textbf{LLaRD} \shortcite{wang2025unleashing} takes LLM to generate preference knowledge and relation knowledge, which are utilized to denoise by maximizing their mutual information.
\end{itemize}


\subsubsection{Case Study of Semantic Relevance and Logical Relevance.}

Here, we provide a case study result to demonstrate the effect of using the LLM-generated Semantic Relevance signals and Logical Relevance signals to identify hard and noisy samples.

\begin{figure}[!htbp]
    \centering
    \includegraphics[width=\linewidth]{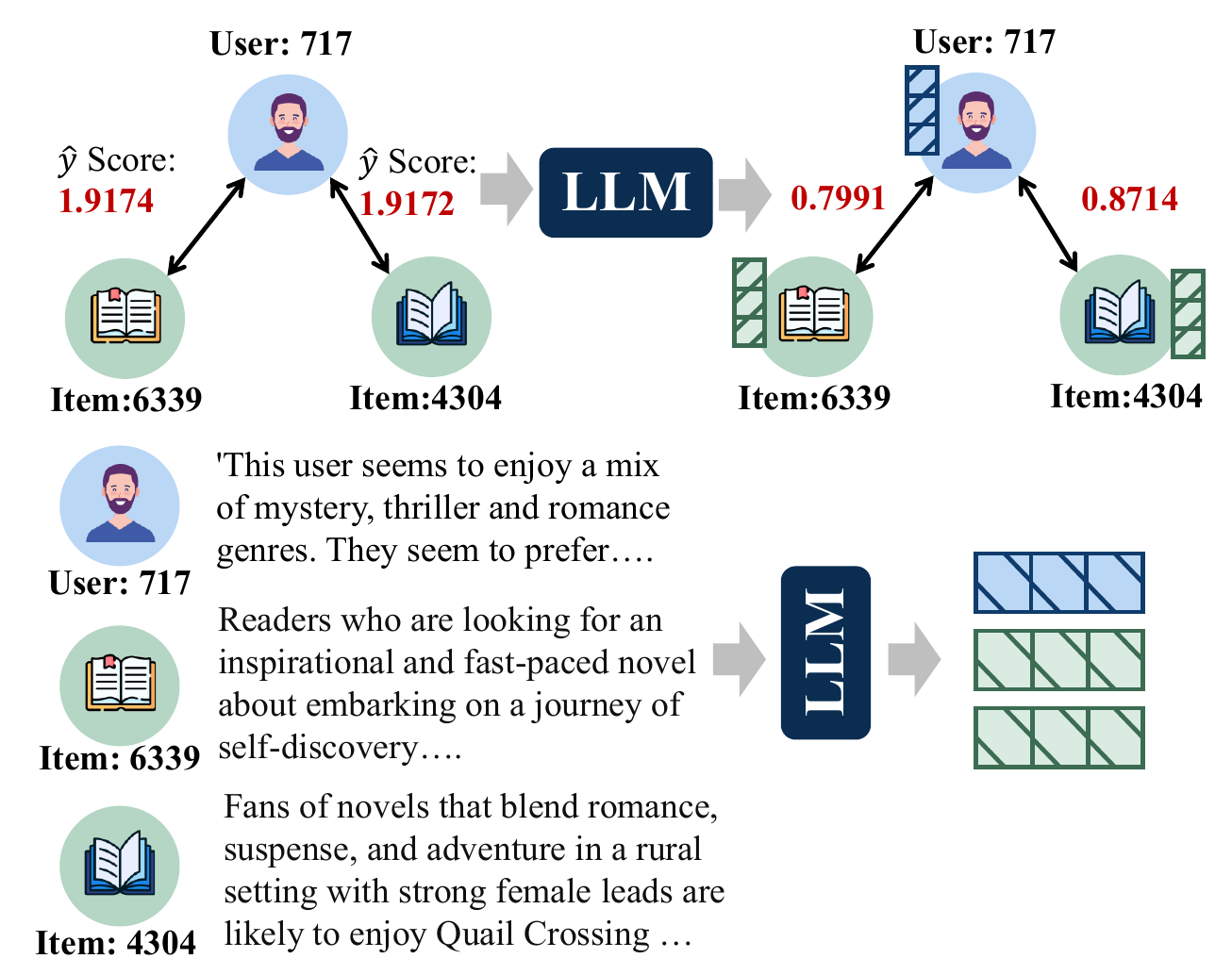}
    \caption{Case study of the Semantic Relevance signal guided sample identification.}
    \label{fig:casesemantic}
\end{figure}

First, we provide an example in Fig. \ref{fig:casesemantic} to show the effectiveness of taking LLM-encoded embeddings to provide semantic relevance signals and identify hard-noisy samples.
As shown in the Fig. \ref{fig:casesemantic}, when the prediction scores of two $(u,i)$ interactions are similar, the LLM-encoded user and item textual embeddings might contain semantic relevance signals that identify hard and noisy sampels with the relatively different embedding similarity scores.
Here, although the user \#717 prefers shows similar prediction scores with the item \#6339 and item \#4304, the LLM-encoded embedding identifies that the user \#717 have a higher similarity score with the item \#4304.
In this way, it provides supplementary information to solve the hard-noisy confusion issue.

\begin{figure}[!htbp]
    \centering
    \includegraphics[width=\linewidth]{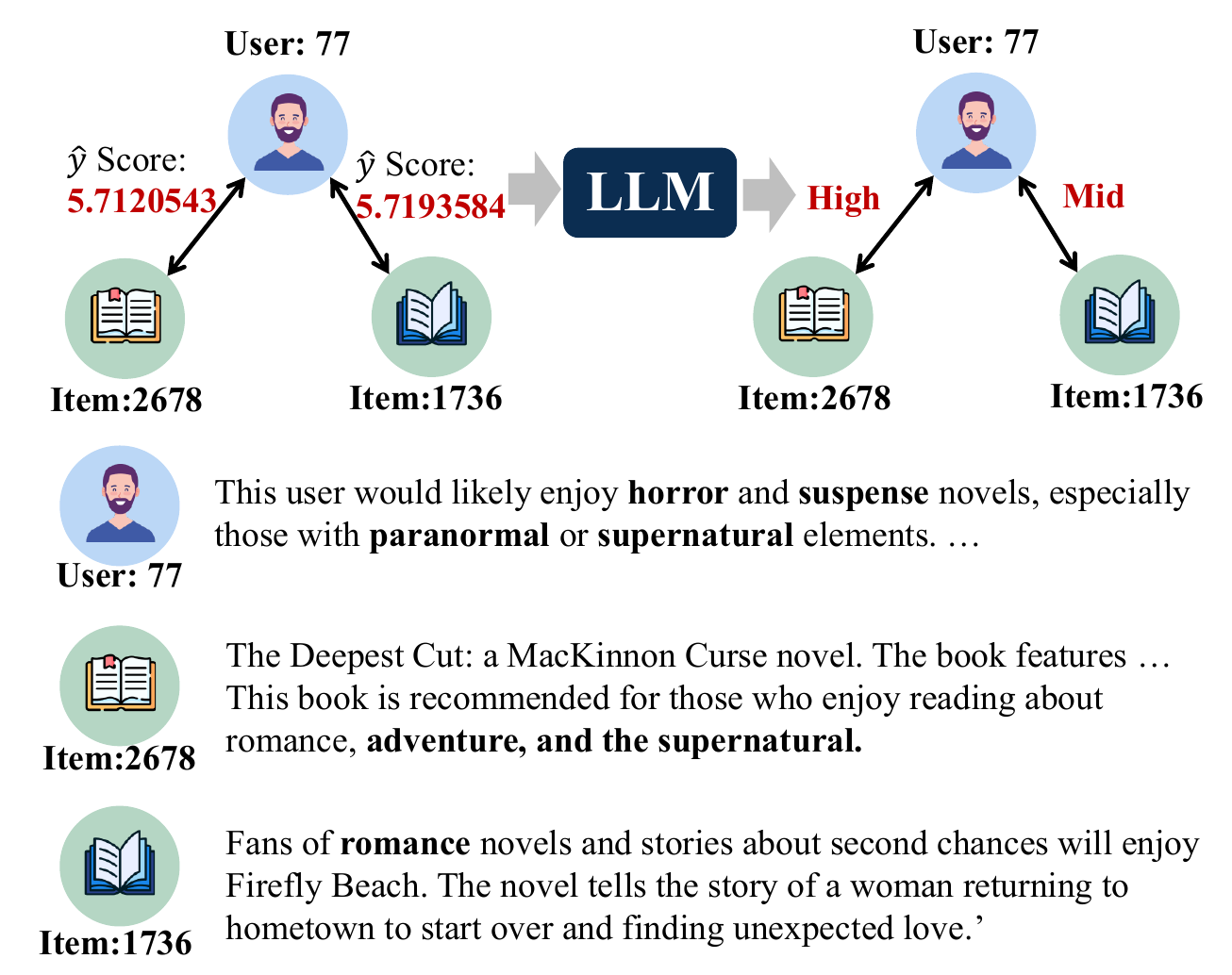}
    \caption{Case study of the Logical Relevance signal guided sample identification.}
    \label{fig:caselogic}
\end{figure}

Second, we demonstrate the effectiveness of LLM-inferred logical relevance in guiding sample identification.
As shown in the Fig. \ref{fig:caselogic}, when the prediction scores of two $(u,i)$ interactions are similar, the LLM can infer the logical relevance of both interactions and differentiate the hard and noisy samples.
Here, as the user \#77 prefers horror and suspense novels with supernatural elements, the LLM analyzes that the user will prefer item \#2678 as it contains supernatural elements, but not item \#1736 as it is more likely to be a classical romance story.

\subsection{Analysis on Embedding Objective Alignment.}

\begin{figure}[!htbp]
    \centering
    \includegraphics[width=\linewidth]{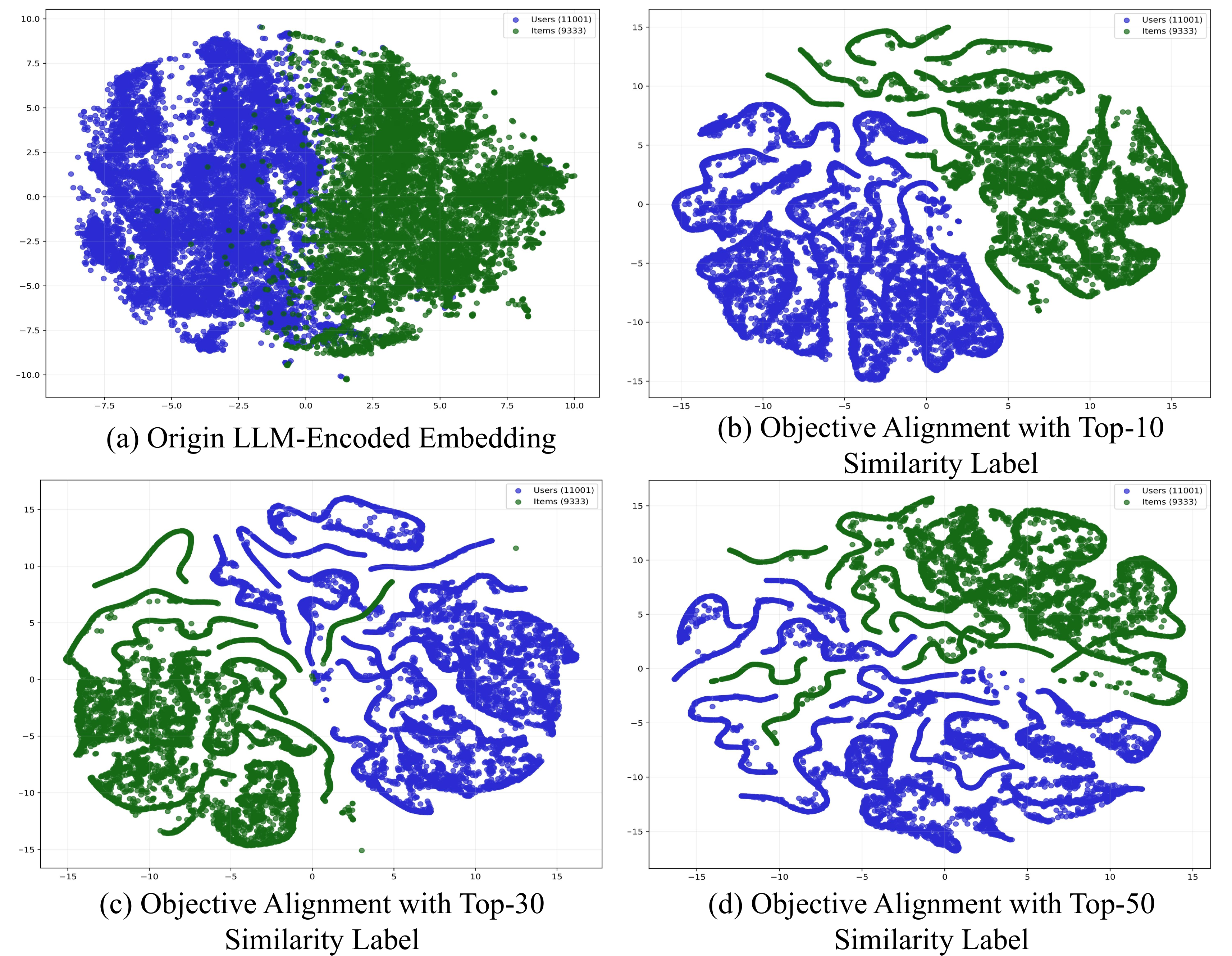}
    \caption{t-SNE visualization of the LLM-encoded user and item embeddings (Blue dots represent the users and Green dots represent the items).}
    \label{fig:objectalign}
\end{figure}

In this section, we provide further analysis of the embedding objective alignment strategy that we proposed in LLMHNI.
More specifically, we analyze the effect of the objective alignment by altering the top-N (N = 10, 30, 50) text embedding similarities threshold utilized in constructing the alignment label. 
We visualize the embedding space with t-SNE as shown in Fig. \ref{fig:objectalign}.
The original LLM-encoded user and item embeddings, as we can see in the Figure \ref{fig:objectalign} (a), are separately distributed in the representation space.
This is a normal phenomenon as the text profiles describing users and items are different in semantics.
However, this does not satisfy the requirements of user-item relevance modeling, which the recommender system model aims to optimize.
From Fig. \ref{fig:objectalign} (b), (c), (d), we can see that by applying the objective alignment strategy, users' and items' representations are gradually becoming intertwined.
This shows the effectiveness of our proposed objective alignment strategy in projecting the original embedding to a task-aligned representation space.
In addition, we also discovered that the larger the $N$, the more $(u,i)$ pairs will be considered as the training label, leading to more intertwined user and item representations.
However, it should also be mentioned that, with too many $(u,i)$ pairs being considered as labels, users' and items' representations are becoming increasingly unevenly distributed.
This poses a great threat to modeling the true semantic relevance between users and items.
It suggests that we should carefully select a suitable top-N threshold when constructing the objective alignment tuning labels.

\subsection{Logical Relevance Analysis Prompt Templates.}
\begin{figure*}[htbp!]
  \centering
  \includegraphics[width=\textwidth]{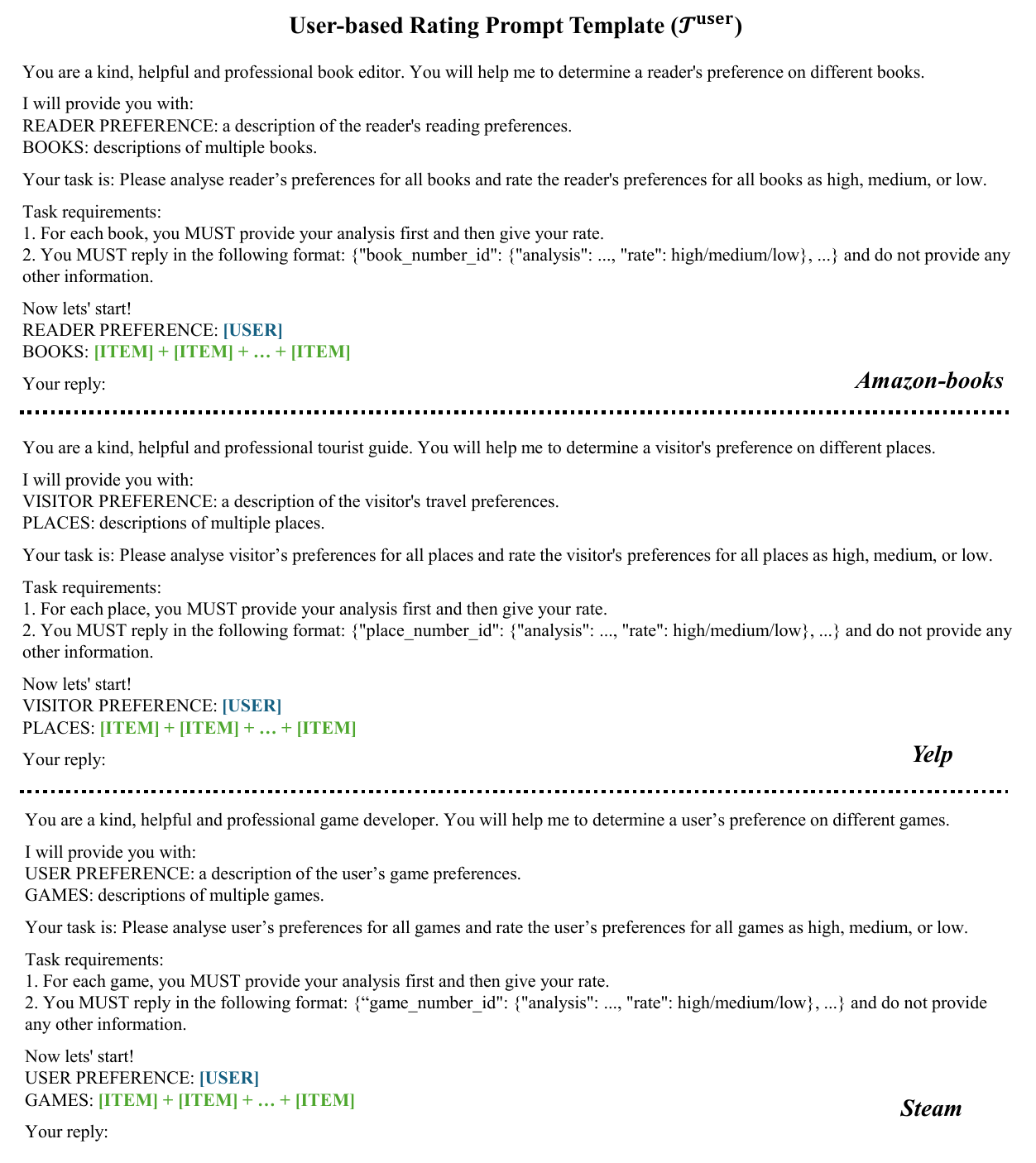}
  \caption{
        Details of user-based rating prompt template on Amazon Books, Yelp, and Steam datasets.
    }
  \label{fig:userbased}
\end{figure*}

\begin{figure*}[htbp!]
  \centering
  \includegraphics[width=\textwidth]{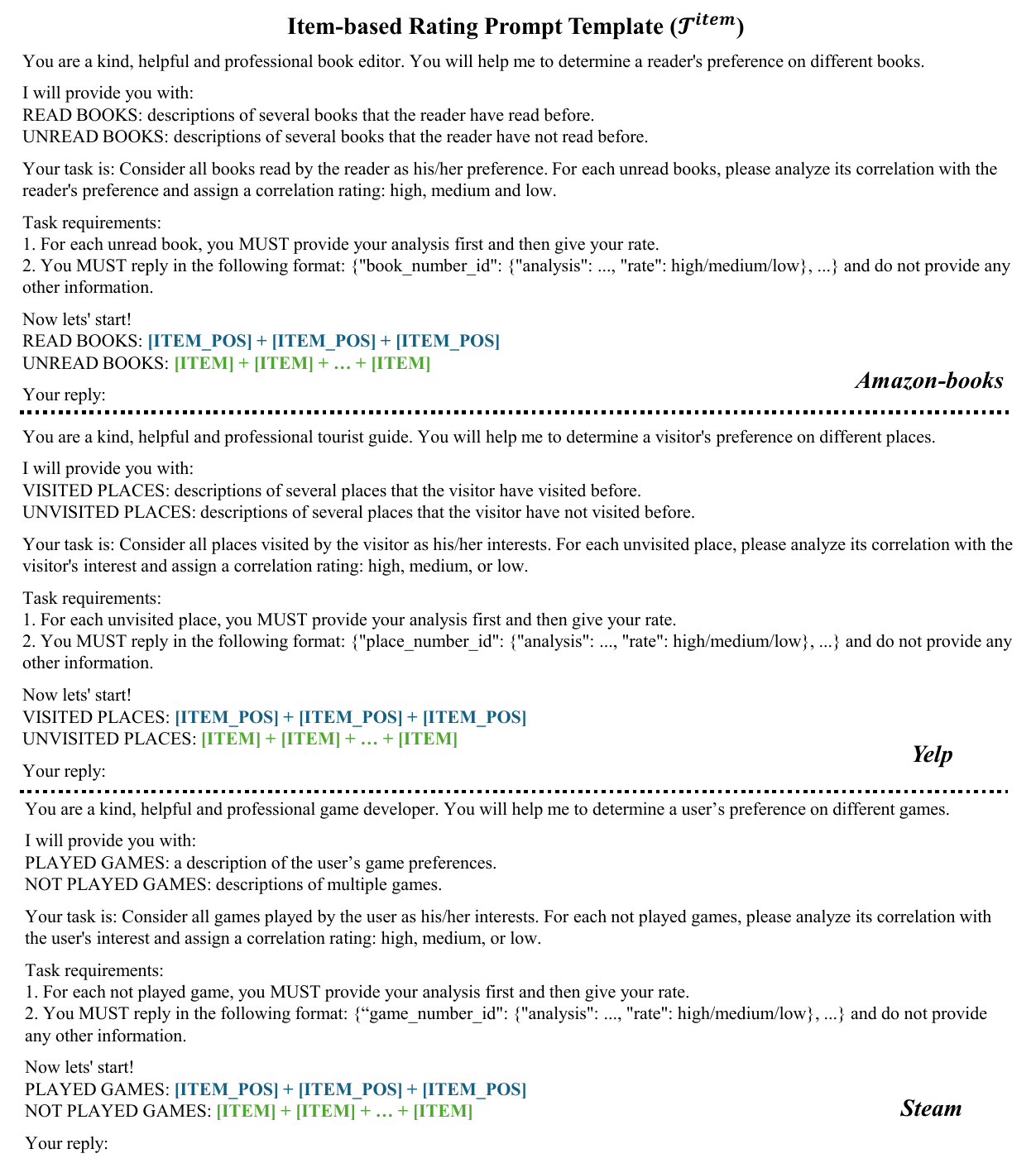}
  \caption{
        Details of item-based rating prompt template on Amazon Books, Yelp, and Steam datasets.
    }
  \label{fig:itembased}
\end{figure*}

In this section, we offer comprehensive information on the templates utilized in the LLMHNI.
We provide the detailed prompt template of user-based rating $\mathcal{T}^{user}$ and item-based sampling $\mathcal{T}^{item}$ for the Amazon-book, Yelp, and Steam datasets.
The user-based preference modeling prompt templates, as shown in Fig. \ref{fig:userbased}, are designed to rate the user's preference on items based on the user preference description in the user profile.
LLM is asked to analyze the logical relevance between the user and item before providing the preference rates.
The user-based preference modeling prompt templates, as shown in Fig. \ref{fig:itembased}, are designed to rate the user preference for items based on the positive items that the user preferred the most.
We also ask LLM to analyze the logical relevance before providing the rates.

\end{document}